\documentstyle[prl,aps,twocolumn,epsfig]{revtex}
\tighten
\def\beq{\begin{equation}}
\def\eeq{\end{equation}}
\def\amu{a_\mu}

\def\gmin2{(g-2)_\mu}

\catcode`\@=11 % This allows one to modify PLAIN macros.

\def\lsim{\mathrel{\mathpalette\@versim<}}
\def\gsim{\mathrel{\mathpalette\@versim>}}
\def\@versim#1#2{\vcenter{\offinterlineskip
    \ialign{$\m@th#1\hfil##\hfil$\crcr#2\crcr\sim\crcr } }}

\def\PL{Phys. Lett.}

\def\PR{Phys. Rev.}
\def\ZPHY{{Z. Phys C} }

\begin{document}
%\begin{flushright}
%{TIFR/TH/01-05}\\
%{CERN-TH/2001-043 }\\
%\date{\today}
%\end{flushright}
%-----------------------------------
\twocolumn[\hsize\textwidth\columnwidth\hsize\csname
@twocolumnfalse\endcsname

\title{Upper Limits on Sparticle Masses from $g-2$ and 
the Possibility for Discovery of SUSY at Colliders and in Dark 
Matter Searches}

\preprint {TIFR/TH/01-05, CERN-TH/2001-043}
\author{Utpal Chattopadhyay}  
\address{Department of Theoretical Physics. Tata Institute of 
Fundamental Research, Homi Bhabha Road, Mumbai 400005, India}

\author{Pran Nath} 
\address{Theoretical Physics Division CERN CH 1211, Geneva, Switzerland\\
Department of Physics, Northeastern University, Boston, MA 02115, USA {\rm (Permanent address)}
}

\vglue 0.2cm

\maketitle

\begin{abstract}
We analyze the implications of the new physics effect seen in the 
$g-2$ Brookhaven  measurement and show that if the effect arises 
from supersymmetry, then the sign of the Higgs mixing parameter 
$\mu$ is determined to be positive in the standard sign convention. 
Further, analyses within the minimal supergravity model
 show that the Brookhaven result leads 
to upper limits on the universal gaugino and scalar masses of 
$m_{\frac{1}{2}}\leq 800$~GeV and $m_0\leq 1.5$~TeV for 
$\tan\beta\leq 55$. Our analysis strongly suggests that supersymmetry 
via production of sparticles must be found at the Large Hadron Collider.
Further, sign($\mu$) positive is favorable for the discovery of supersymmetric 
cold dark matter. 

\vglue 0.4cm


\end{abstract}
\vglue 0.4cm
]

The anomalous magnetic moment of the muon $ a_{\mu}=(g_{\mu}-2)/2$ 
(where g relates the magnetic moment $\vec \mu $ of a particle to its 
spin $\vec S$ by $\vec \mu =g(e/2m)\vec S$)
is one the most accurately determined quantities in physics.
 Recently, the experiment E821
at the Brookhaven National Laboratory has made an more
 precise determination of 
$a_{\mu}$\cite{brown}.
 The new measurement is in good agreement with the previous 
 determinations but the combined error is now reduced
 by a factor of about 3.  The  world average of the experimental
 results including the new measurements is given by
 $a_{\mu}^{exp}=11659203(15)\times 10^{-10}$.
 The Standard Model prediction for  $a_{\mu}$ including up to 
$\alpha^5$ QED corrections, $\alpha^2$ and $\alpha^3$ hadronic 
corrections and up to two loop electro-weak corrections gives\cite{marciano}
$a_{\mu}^{SM}=11659159.6(6.7)\times 10^{-10}$ where essentially the 
entire error in the Standard Model prediction comes from the 
error in the hadronic  corrections\cite{davier}.
 Remarkably the new Brookhaven measurement measurement 
finds a 2.6 sigma difference between the experiment 
and the Standard Model result\cite{marciano} signaling the
 possible onset of new physics\cite{brown}, i.e., 

\beq
a_{\mu}^{exp}-a_{\mu}^{SM}=43(16)\times10^{-10}
\eeq
It has been known for some time that 
$a_{\mu}$ is sensitive to new physics such as 
supersymmetry (SUSY)\cite{fayet,kosower,yuan}.
Specifically, estimates of the correction in the
well motivated  supergravity unified (SUGRA) model  
showed  in 1983-1984 that the supersymmetric correction to $a_{\mu}$ 
can be as large or larger\cite{yuan} than  the 
Standard Model electro-weak correction\cite{marciano,fuji}.
The more recent analyses\cite{lopez,chatto,cgr} 
support the previous conclusions\cite{yuan} 
that the supersymmetric electro-weak
effects can be large. Further, it is known that large CP effects
can be consistent with the electron and the neutron electric dipole moment 
constraints\cite{inedm} and analyses show that the CP violations can
generate large corrections to $a_{\mu}$\cite{in}.
A variety of other effects such as arising 
from extra dimensions, anomalous W-couplings etc. have
also been examined (for a review see Ref.\cite{marciano2}).   
In the following we analyze the implications
of the new result from Brookhaven on sparticle masses. 
First, we show that the Brookhaven experiment determines
the sign of the Higgs mixing parameter $\mu$. 
We then analyze the limits on
the sparticle spectrum by using the constraint of Eq.(1), 
 taking a 2$\sigma$ error corridor and attributing the entire 
 difference between theory and experiment to supersymmetry, which
 gives
$10.6\times 10^{-10} <a_{\mu}^{SUSY}<76.2\times 10^{-10}$,
where $a_{\mu}^{SUSY}=a_{\mu}^{exp} -a_{\mu}^{SM}$.
For the purpose of definiteness we shall give an analysis of the 
constraint mostly within the framework of SUGRA  model
and specifically in its minimal form (mSUGRA)\cite{sugra}.
However, for comparison we also discuss the results within
the minimal anomaly mediated supersymmetry breaking (AMSB) 
scenario following the analysis of Ref.\cite{cgr}.
 At low energy mSUGRA can be parameterized by 
  $m_0, m_{\frac{1}{2}}, A_0, tan\beta, sgn({\mu})$ where
  $m_0$ is the universal scalar mass, $m_{\frac{1}{2}}$ is the
  universal gaugino mass,  
  $A_0$ is the universal trilinear coupling at the grand unified theory 
  scale and $\tan\beta =<H_2>/<H_1>$ where $<H_2>$ gives mass to the up 
  quark and $<H_1>$ gives mass to the down quark and the lepton,
  and $\mu$ is the Higgs mixing parameter which appears in the 
  superpotential in the form $W^{(2)}=\mu H_1H_2$ (Our sign convention on 
  $\mu$ is that of Ref.\cite{sugraworking}).
 The supersymmetric contributions $a_\mu^{SUSY}$
at the one loop level consists
of the chargino-sneutrino exchange and  of the neutralino-smuon 
exchange so that 
$a_\mu^{SUSY}=a_\mu^{\tilde \chi^\pm}+a_\mu^{\tilde \chi^0}$.
However, typically it is the chargino-sneutrino exchange 
contribution $a_\mu^{\tilde \chi^\pm}$ that dominates. 

 It was noted in two previous papers several years ago 
 (see Ref.\cite{lopez} and  
 Chattopadhyay and Nath (CN)in Ref\cite{chatto}) that the
 sign of $\amu^{SUSY}$ is correlated with the sign of $\mu$.
 It was shown by CN in Ref.\cite{chatto} that this
 correlation arises because of the signature carried by the
 contribution of the light chargino exchange term in
  the chiral left and the chiral right interference term in the chargino
  exchange contribution. It was found that over most of the 
  parameter space one has
 $ \amu^{SUSY}>0$ for $\mu>0$ and $\amu^{SUSY}<0$ and  for $\mu<0$, 
except when $\tan\beta$ is very close to 1. 
Since the data from the BNL experiment indicates $\amu^{SUSY}>0$
 we conclude that
 \begin{equation}
 \mu >0 ~~~~~~(from ~~BNL ~~data\cite{brown})
 \end{equation}
The fact that the sign of $\mu$ is determined by the BNL data turns 
out to be positive
is of great consequence. It is known that the current
experimental limits on the flavor changing neutral current 
process $b \rightarrow s\gamma$
eliminates  a majority of the parameter space for 
$\mu<0$\cite{bsgamma} and consequently the neutralino-proton 
cross-sections ($\sigma_{\chi^0-p}$)for the direct detection of dark matter are
significantly smaller.  Current direct search experiments
are  sensitive to $\sigma_{\chi^0-p}\geq 1\times 10^{-6}$pb and 
one expects future experiments to reach a sensitivity of 
$\sigma_{\chi^0-p}\geq 1\times (10^{-9}-10^{-10})$pb and this sensitivity
can probe a majority of the parameter space of mSUGRA for $\mu>0$.
However, this is not the case for $\mu<0$\cite{santoso}.
Thus a positive $\mu$ sign given by Eq.(2) using the BNL data is very
encouraging for the discovery of neutralino cold dark 
matter\cite{bsgamma,santoso}.

We discuss now the other consequences of the BNL $g-2$ constraint.
In Fig.1 we plot the allowed region in the sneutrino and the light
chargino plane for $\mu>0$ for values of $\tan\beta$ of 5, 10, 30, 45 and 
55.   
We find the remarkable result that one has now an upper limit
on the chargino and sneutrino masses. Thus one finds that the sneutrino mass
lies below 1.1 TeV and the chargino mass lies  below 590 GeV
for $\tan\beta = 30$, the sneutrino mass
lies below 1.4 TeV and the chargino mass  lies  below 650 GeV
for $\tan\beta = 45$, and 
the sneutrino mass
lies below 1.5 TeV and the chargino mass  lies  below 500 GeV
for $\tan\beta = 55$.  
 A similar situation occurs in the $m_0-m_{\frac{1}{2}}$ plane.
 In Fig.2 we give a plot of  the allowed region in the $m_0-m_{1/2}$
 plane for $\tan\beta$ of 5,10, 30, 45 and 55. Here one finds an upper limit on $m_0$ of 1.1 TeV and on 
 $m_{\frac{1}{2}}$ of 750 GeV for $\tan\beta = 30$,
 an upper limit on $m_0$ of 1.4 TeV and on 
 $m_{\frac{1}{2}}$ of 800 GeV for $\tan\beta = 45$, 
 and 
  an upper limit on $m_0$ of 1.5 TeV and on 
 $m_{\frac{1}{2}}$ of 625 GeV for $\tan\beta = 55$. 
  There are also
  interesting lower limits on the parameters in Figs.1 and 2. 
 Additionally, one finds that the allowed parameter space accommodates
 a light Higgs of 115 GeV which is in the vicinity of the lower 
 limit on the Higgs mass from the CERN Large Electron-Positron Collider 
data.  We also note from Figs.(2a) and Fig.(2b) that $m_0\leq 550$ GeV 
$m_{\frac{1}{2}}\leq 475$ GeV  for $\tan\beta \leq 10$.
The lower limits, e.g. of the light chargino mass, in this case 
are constrained only by experiment which for the light chargino
is about 100 GeV. Thus in this part of the parameter space the
light chargino could be accessible at the Fermilab collider Tevatron Run 2 
(RUNII) since RUNII will be able to explore chargino masses up
to about 200 GeV via the trileptonic signal.
   Thus the implication is that if  
 $\tan\beta \sim 5-10$,  it would be much easier to discover sparticles
 at accelerators. However, there is no guarantee that we shall find
 ourselves in this region of the parameter space and thus the
 discovery of sparticles though possible at the Tevatron is not guaranteed.
  Returning to the upper limits we can say that under
 the constraint of the BNL $g-2$ experiment 
 we find that
 \begin{equation}
m_{\frac{1}{2}}\leq 800~GeV, m_0\leq 1.5 TeV
(\tan\beta\leq 55)
 \end{equation}
 Although more detailed mapping of the parameter space would 
 modify the limits somewhat, we expect the main result to survive 
 with some modest corrections. 
However, one must take the result of Eq.(3) with  caution. 
We  note that Eq.(1) exhibits
a new physics effect at the $2.6 \sigma$ level and there were 4 times
more data collected by the BNL group in the year 2000 which will be 
analyzed in the 
near future, and an important further check of the result will occur.
To check if our upper limits of Eq.(3) are robust 
we have also analyzed the $g-2$ constraint within the minimal AMSB scenario
following the analysis of Ref.\cite{cgr} to which the reader is referred
to for details and references. 
The results are presented in Fig.3 for $\mu>0$ and 
$\tan\beta =10,30,40$ ($\tan\beta=45$ in this case is not allowed because
of the CP odd Higgs boson turning tachyonic which in turn results 
from a large b-quark SUSY QCD correction~\cite{cgr}).  
Here one finds an upper limit on the chargino mass 
of about 300 GeV and on the sneutrino mass  of 
about 1.1 TeV. These limits are lower than those implied by 
Eq.(3). Thus the upper limits of Eq.(3) are very robust.
Eq.(3) implies an upper limit on the light chargino mass of about
650 GeV, and upper limits for the  gluino mass and squark masses of
about 2 TeV. Since the CERN Large Hadron Collider (LHC) 
will be able to see the squarks and gluinos till about
 2 TeV\cite{cms}, the above results provide a strong evidence 
for the possibility that  
sparticles must become visible at the LHC.
 Further, an analysis of the effects of extra dimensions from Kaluza-Klein
 excitations  on $g-2$  show that these
 do not provide a strong background to SUSY\cite{ny2}.

 In conclusion, the BNL data provides a determination of the 
 sign of the Higgs mixing parameter and we find this sign 
 to be positive in the standard notation\cite{sugraworking}. 
 Further, assuming that the
 entire difference between experiment and the Standard Model
 result comes from SUSY, we find
 that the result from the Brookhaven experiment implies an upper 
 bound on the SUSY parameters
 $m_0$ and $m_{\frac{1}{2}}$ which we find to lie in a region 
 accessible to future accelerators. Thus we conclude that the
 major implication of the BNL $g-2$ result is that sparticles must  
 become visible  at the LHC and possibly at RUNII of the 
 Tevatron. We also find that the parameter space allowed by
 the $g-2$ experimental limits allow a Higgs mass 
 of 115 GeV which is in the vicinity of the lower limit from LEP.
 The positive $\mu$ sign and the upper limits on the
 sparticles masses implied by the BNL $g-2$ data is also encouraging
 for the discovery of neutralino dark matter in dark matter 
 searches. In the analysis above we did not impose the 
 $b\rightarrow s+\gamma$ constraint
 and the relic density constraint. These are more model dependent and would
 tend to only reduce the upper limits of Eq.(3). Thus our upper limits
 of Eq.(3) and our prediction of the observation of sparticle at the
 LHC are robust.

This research was supported in part by NSF grant PHY-9901057.

\begin{figure}[hbt]
\centerline{\epsfig{file=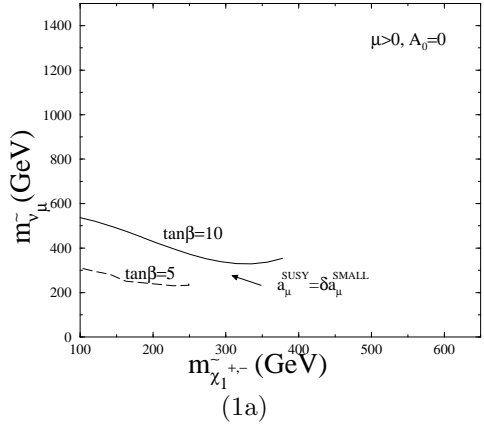,width=0.35\textwidth}}
%\vspace*{-1.0in}
%\end{figure}
%\begin{figure}[hbt]
\vspace*{-0.20in}
\begin{center}
(1a)
\end{center}
\centerline{\epsfig{file=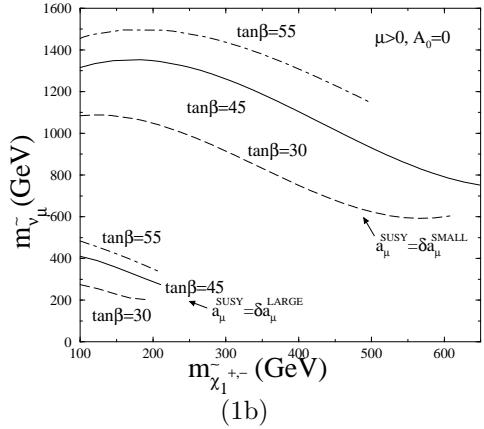,width=0.35\textwidth}}
\vspace*{-0.20in}
\begin{center}
(1b)
\end{center}
\caption{
A plot (a) of the upper limits in the sneutrino-light 
chargino plane for mSUGRA from the lower limit of $a_{\mu}^{SUSY}$
when $\tan\beta$ = 5 and 10, $A_0=0$ and $\mu>0$. Similar plots (b) 
for $\tan\beta$ = 30, 45 and 55 have both upper and lower limits. The
$g-2$ bounds are $\delta a_\mu^{\rm SMALL}=10.6 \times 10^{-10}$, and
$\delta a_\mu^{\rm LARGE}=76.2 \times 10^{-10}$
}
\end{figure}
\begin{figure}[hbt]
\centerline{\epsfig{file=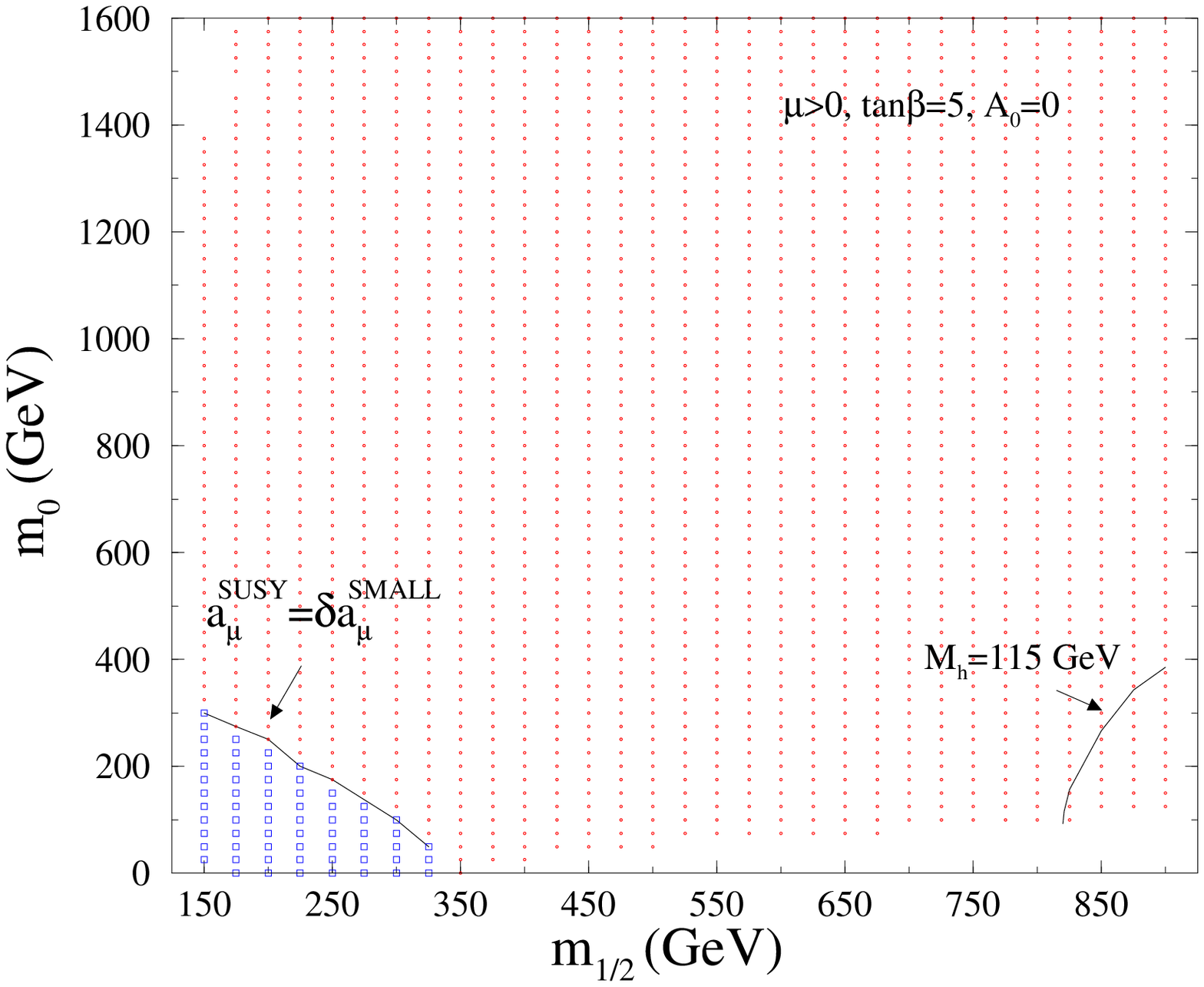,width=0.35\textwidth}}
\vspace*{-0.20in}
\begin{center}
(2a)
\end{center}
\centerline{\epsfig{file=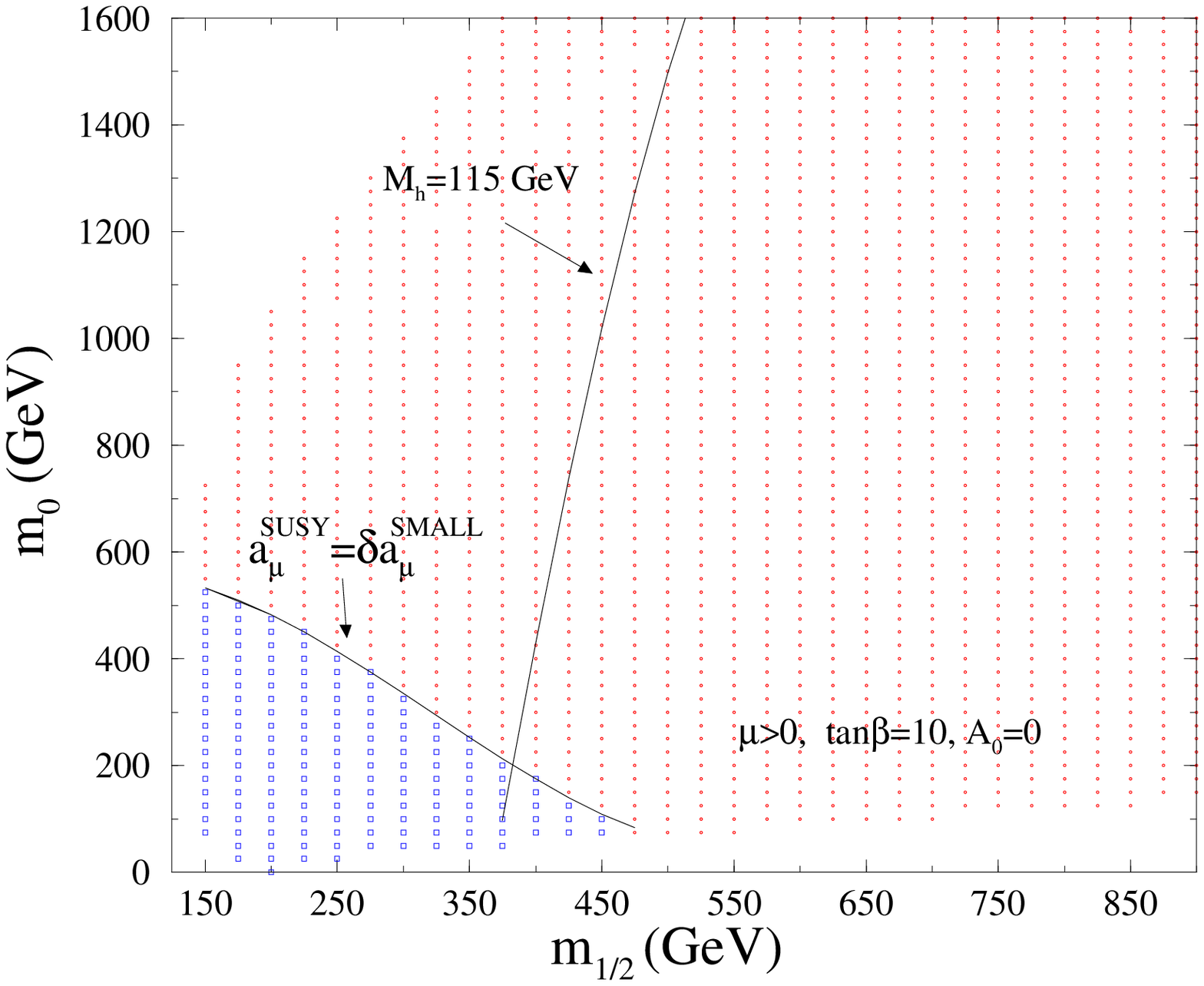,width=0.35\textwidth}}
\vspace*{-0.20in}
\begin{center}
(2b)
\end{center}
\centerline{\epsfig{file=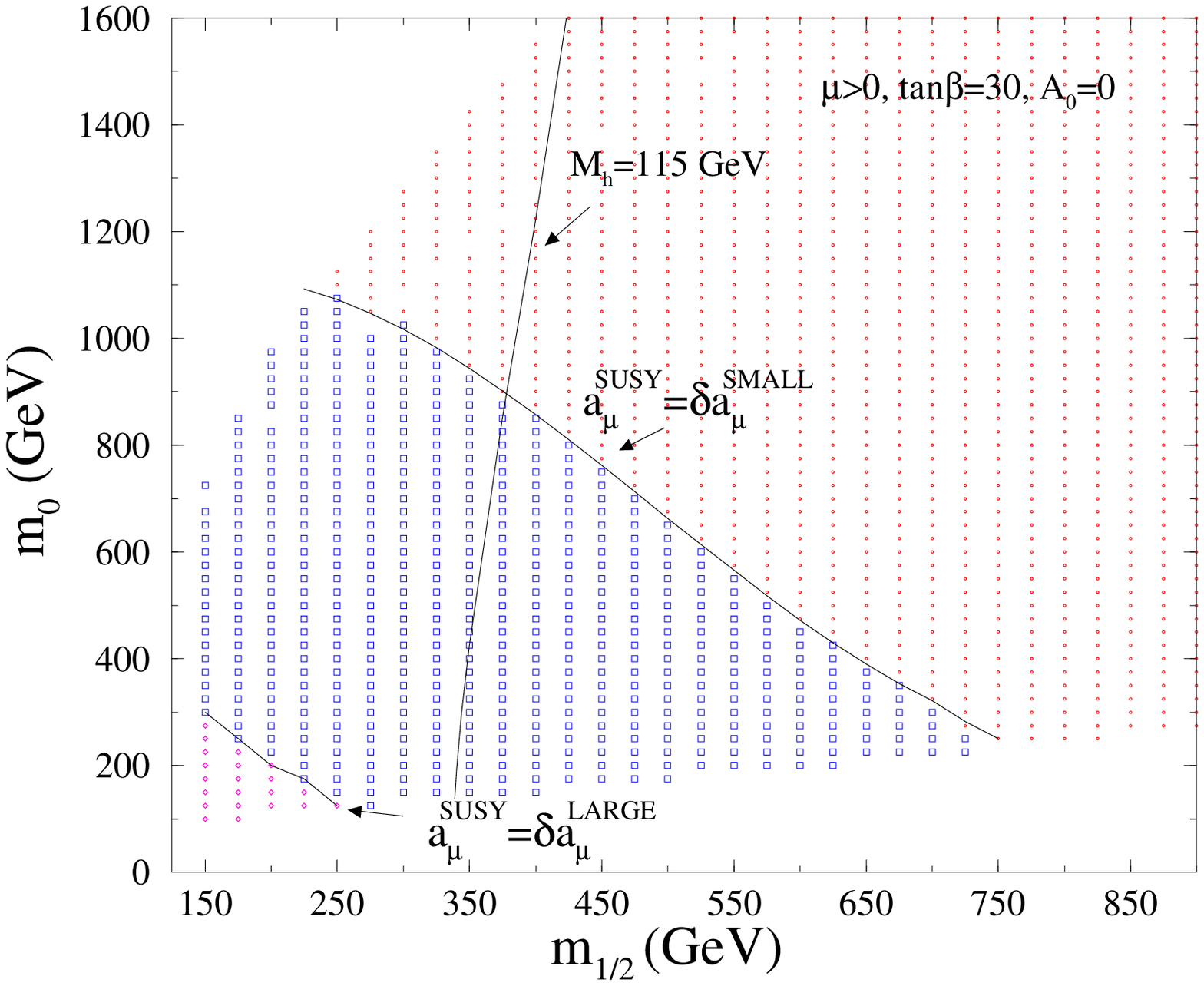,width=0.35\textwidth}}
\vspace*{-0.20in}
\begin{center}
(2c)
\end{center}
\centerline{\epsfig{file=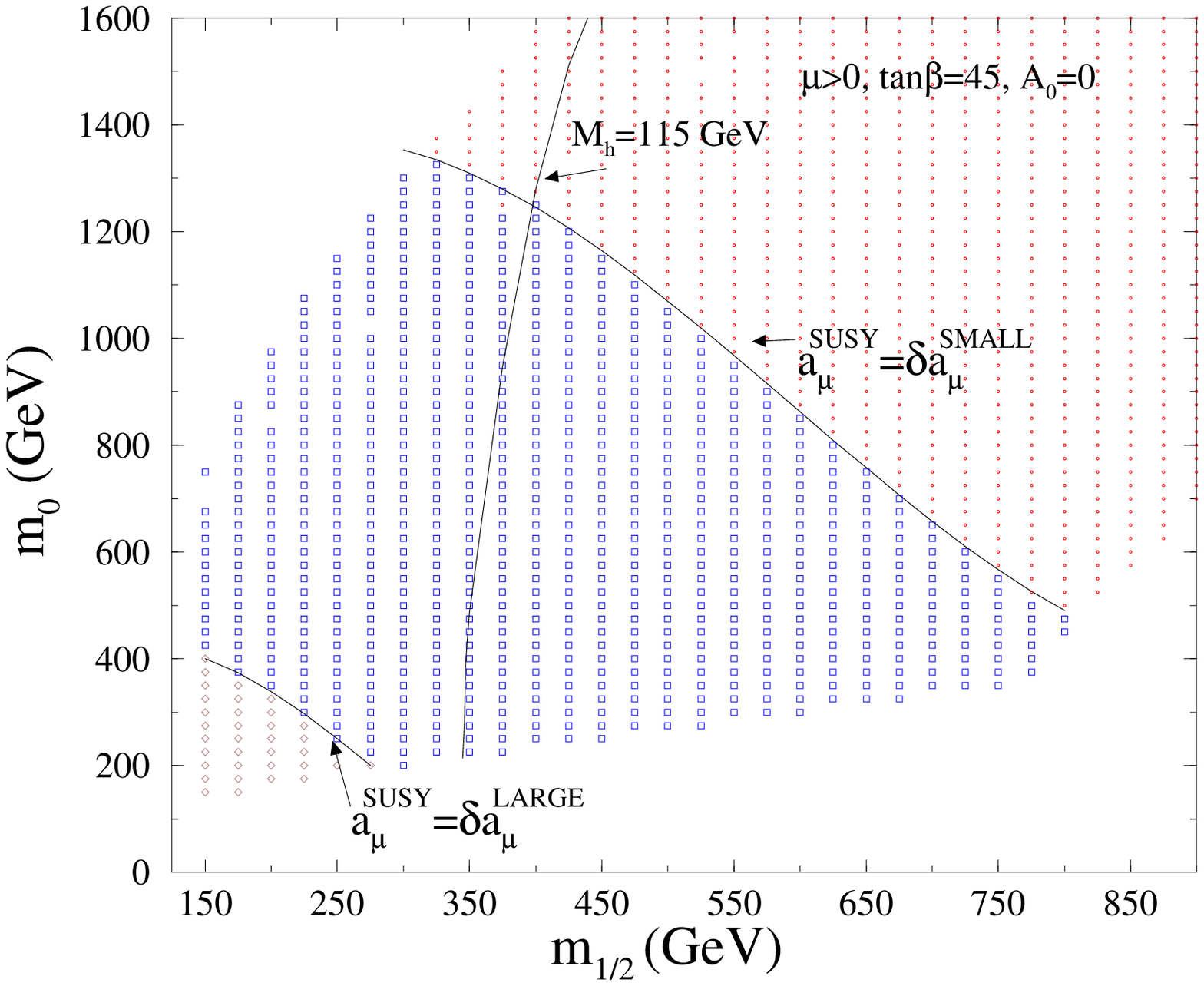,width=0.35\textwidth}}
\vspace*{-0.20in}
\begin{center}
(2d)
\end{center}
\centerline{\epsfig{file=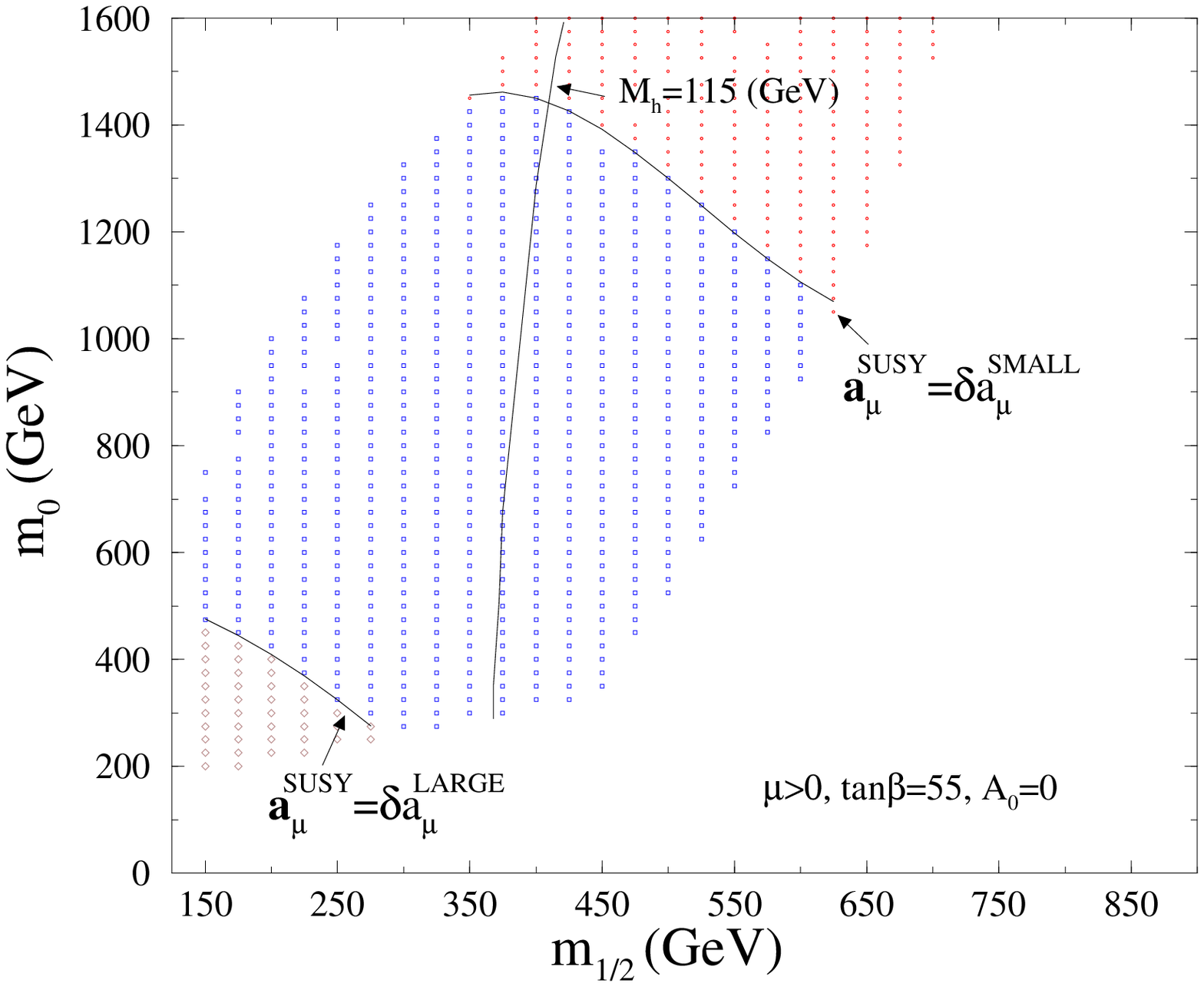,width=0.35\textwidth}}
\vspace*{-0.20in}
\begin{center}
(2e)
\end{center}
\caption{
 Plots of the regions allowed (blue squares) by the $g-2$ constraint in the 
$m_0-m_{\frac{1}{2}}$ plane in mSUGRA when $\tan\beta=$5, 10, 30, 45 and 55, 
with  $A_0=0$ and $\mu>0$.
}
\end{figure}
\begin{figure}[hbt]
\centerline{\epsfig{file=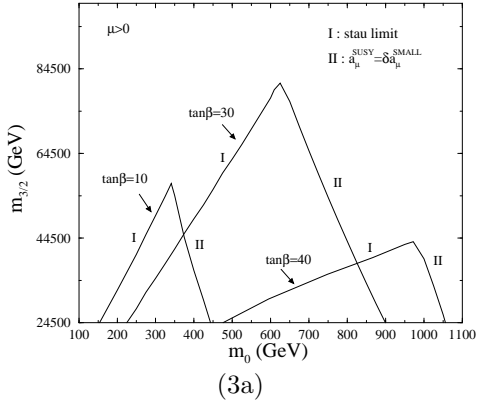,width=0.35\textwidth}}
\vspace*{-0.20in}
\begin{center}
(3a)
\end{center}
\centerline{\epsfig{file=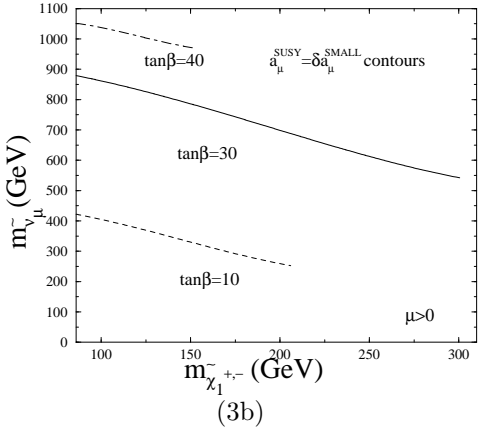,width=0.35\textwidth}}
\vspace*{-0.20in}
\begin{center}
(3b)
\end{center}
\caption{
(a)
A plot of the region allowed (inside of the triagular areas)
by the $g-2$ constraint 
in the $m_{\frac{3}{2}}-m_0$ plane  for the minimal AMSB model 
when $\tan\beta=$ 10,30, and 40. The regions above the left 
side arm of each triangle are disallowed due to stau turning
tachyonic; (b) same allowed regions when plotted in the
sneutrino-light chargino plane.
}
\end{figure}

\end{document}